\documentclass[]{interact}

\usepackage{epstopdf}
\usepackage[caption=false]{subfig}

\usepackage[numbers,sort&compress,merge]{natbib}
\usepackage{xcolor}
\bibpunct[, ]{[}{]}{,}{n}{,}{,}
\renewcommand\bibfont{\fontsize{10}{12}\selectfont}

\theoremstyle{plain}

\theoremstyle{definition}

\theoremstyle{remark}

\begin{document}



\title{Vapour-Liquid equilibrium and low-temperature liquid-crystal phase diagram of discotic colloids}

\author{
\name{Alejandro Cuetos\textsuperscript{a}*, Bruno Mart\'{\i}nez-Haya\textsuperscript{a} and Jos\'e Manuel Romero-Enrique\textsuperscript{b,c}}
\affil{
\textsuperscript{a} Center for Nanoscience and Sustainable Technologies (CNATS). Department of Physical, Chemical and Natural Systems, Universidad Pablo de Olavide, 41013 Seville, Spain\\
\textsuperscript{b}Departamento de F\'{\i}sica At\'omica, Molecular y
Nuclear, \'Area de F\'{\i}sica Te\'orica, Universidad de Sevilla,
Avenida de Reina Mercedes s/n, 41012 Seville, Spain\\ \textsuperscript{c}Instituto Carlos I de F\'{\i}sica Te\'orica y Computacional, Campus Universitario Fuentenueva,
Calle Dr. Severo Ochoa, 18071 Granada, Spain\\
\textsuperscript{*}Author for correspondence: acuemen@upo.es}
}

\maketitle

\begin{abstract}
Discotic colloids give rise to a paradigmatic family of liquid crystals with sound applications in Materials Science. 
In this paper, Monte Carlo simulations are employed to characterize the low-temperature liquid crystal phase diagram and the vapour-liquid coexistence of discotic colloids interacting via a Kihara potential. Discoidal particles with thickness-diameter aspect ratios $L^*\equiv L/D$=\,0.5, 0.3, 0.2 and 0.1 are considered. For the less anisotropic particles ($L^*$$\ge$0.2), coexistence of a vapour phase with the isotropic fluid and with the columnar liquid crystal phase is observed. As the particle anisotropy increases, the vapour-liquid coexistence shifts to lower temperatures and its density range diminishes, eventually merging with coexistences involving the liquid crystal phases. The $L^*=$\,0.1 fluid displays a rich sequence of mesophases, including a nematic phase and a novel lamellar phase in which particles arrange in layers perpendicular to the nematic director.  
\end{abstract}


\section{Introduction}
Liquid crystals show a plethora of mesophases in which there is long-ranged orientational ordering in addition to partial spatial ordering in some directions \cite{DeGennes1993}. The characteristics of the phase diagram are  determined to a large extent by anisotropy of the molecular repulsive core. In this way, calamitic rod-like mesogens typically show nematic and smectic phases, while discotic mesogens display nematic and columnar phases. Intermolecular interaction features can affect the phase diagram as a result of the interplay between repulsive and attractive contributions from dispersion forces \cite{deMiguel92,VEG92,deMiguel1,WU} or multipolar electrostatic interactions \cite{Levesque,Williamson,McGrother,Houssa,Houssa2,Williamson2,Houssa3,Tre23}. For example, a transition between isotropic phases characterized by low density (vapour) and moderate density (liquid) can be observed only if the intermolecular potential has an attractive contribution. 

This paper explores the phase diagram of discotic particles in the low-temperature region, devoting special attention to the appearance of the vapour-liquid transition and its interplay with the different mesophases that the system may present. Discotic particles may mimic molecular species in thermotropic liquid crystals \cite{BIS10,WOH16}, which can be modelled via a full-atom model or, alternatively, by coarse-grained models. The latter may also be used to study colloidal lyotropic liquid crystals, such as aluminum hydroxide gibbsite platelets and nontronite or laponite mineral clays \cite{Brown2,vanderKooij,vanderKooij2,vanderKooij3,Liu,Fossum,Michot}. Some examples of coarsed-grained models of discotic molecules considered in the literature are disk-like hard-core models \cite{Veerman}, the Gay-Berne model \cite{GB}, the Kihara model \cite{KIH53,Gam08} or the combined Gay-Berne-Kihara model \cite{MAR09,MOR21}.

From a computer simulation perspective, most studies on phase diagrams of mesogens have focused on the identification and classification of the type of stable mesophases. For the Gay-Berne model, only a few consider the vapour-liquid (either isotropic or nematic) phase transition, as well as its dependence on the potential parameters, i.e. the length-to-breath ratio of the particle $\kappa$ and the energy depth anisotropy $\kappa'$. For calamitic mesogens, refs. \cite{deMiguel,Elvira,deMiguel1} report the effect of the attractions on the vapour-liquid transition by using Gibbs-ensemble Monte Carlo simulations, as well as direct coexistence techniques. For the $\kappa$=\,3 fluid, evidence of vapour-liquid coexistence was found for different values of $\kappa'$=\,1--5 \cite{deMiguel,Elvira,deMiguel1}. Moreover, for a fixed value of $\kappa'$=\,5, ref. \cite{Brown} reports that a vapour-liquid transition is only observed for $\kappa<$\,4. Regarding oblate Gay-Berne molecules, most of the simulation studies focus on the nematic-columnar phase transition \cite{Emerson2,Bates4,Ryckaert1,Ryckaert2,Chakrabarti} . The liquid-vapour coexistence has been investigated with direct coexistence and zero-pressure procedures \cite{Frenkel} for models with $\kappa$=\,0.3 and $\kappa$=\,0.5 \cite{RUL17a,RUL17b}. 

This paper focuses on discotic particles modelled via the Kihara interaction potential. The vapour-liquid coexistence of this model has been studied for both prolate and oblate particles at comparably higher temperatures than the ones considered here \cite{Gam08,Tre23}. The effect of dipolar and quadrupolar interactions in this transition has also been assessed for discotic particles \cite{Tre23}. Moreover the liquid-crystalline phase diagram of the Kihara fluis has been explored for rod-like \cite{VEG91,VEG92,CUE03} and disk-like \cite{MAR09} particles. Also worth mentioning, a modified version of the Kihara model which introduces orientationally-dependent attractive interactions in the spirit of the Gay-Berne model exposed a particularly rich liquid crystal phase diagram \cite{MOR21}.

The paper is organized as follows. Section 2 is devoted to the description of the colloidal model and the simulation techniques used in this work. The analysis of the vapour-liquid coexistence curves and their interplay with the different observed mesophases is done for different colloidal aspect ratios in Section 3. The paper closes with the main conclusions of our work.

\section{Simulation Methods}

\begin{figure}[!ht]
	\centering
	\hspace*{-0.3cm}
	\includegraphics[width=9cm]{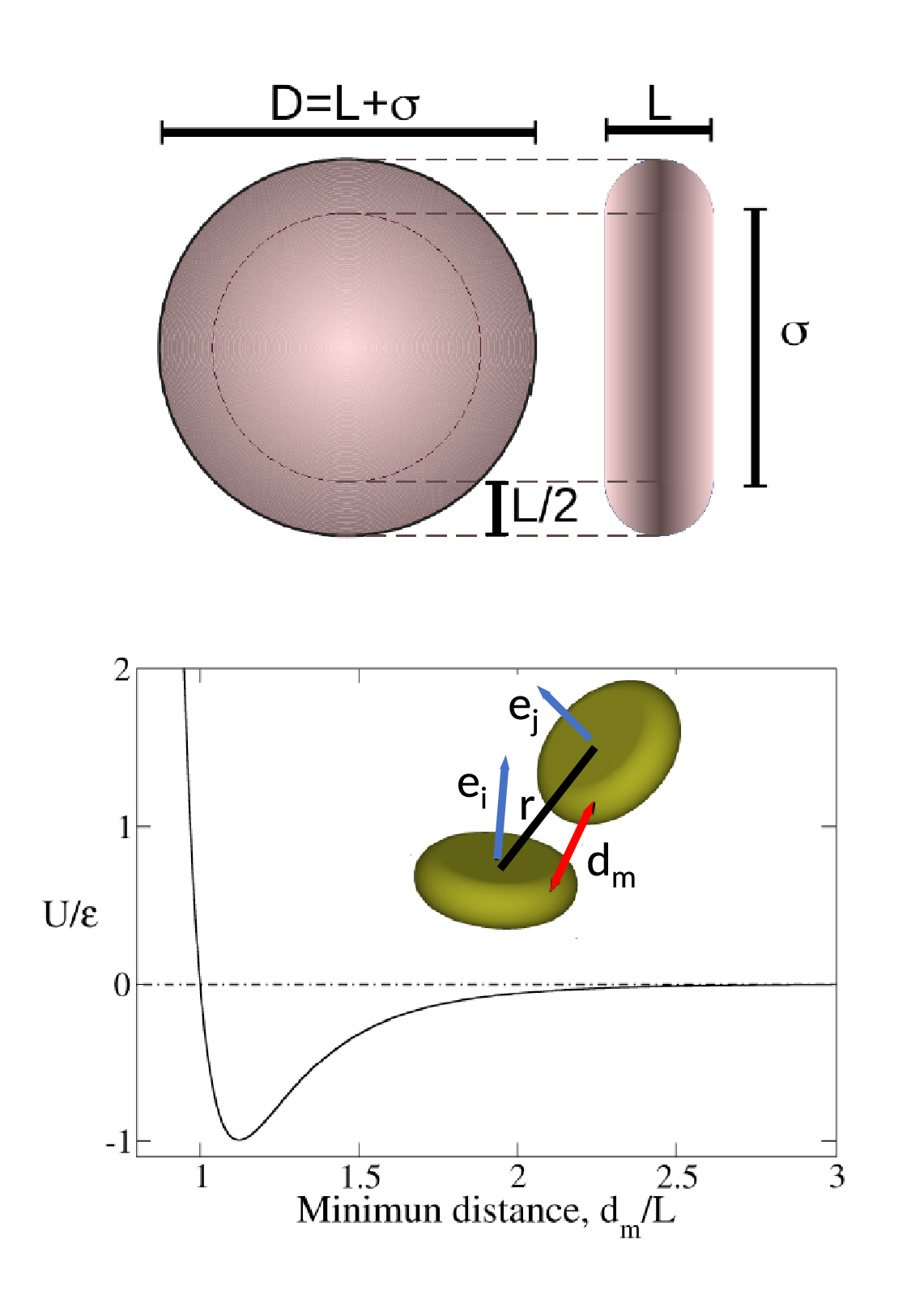}
	\vspace*{-0.4cm}
	\caption{Top: Main geometric parameters of the spherocylindrical particles associated with the discotic Kihara model (particle width L and diameter D, and diameter of the core disk $\sigma$). Bottom: Lennard-Jones-type dependence of the Kihara interaction potential on the minimum distance between particles, $d_m$. The inset shows the positional and orientational vectors involved in the calculation of the minimum distance.}
	\label{fig1}
\end{figure}

\subsection{Colloidal model}

The low-temperature phase behaviour of discoidal colloids is investigated here within the coarse-grained Kihara interaction model. This model assigns an oblate spherocylindrical shape to the colloidal particles \cite{BOU86}. A spherocylinder can be formally viewed as a body of revolution generated by rotating a rectangle capped by two semicircles, as illustrated in Fig\,\ref{fig1}. Depending on the axis of revolution, an oblate (disk-like) or a prolate (rod-like) spherocylinder may be generated. For a discoidal particle built from a rectangle with long axes $\sigma$ and short axes $L$, the resulting spherocylinder has a total diameter of $D$=\,$L$+$\sigma$ and the thickness/diameter aspect ratio is given by $L^*$=\,$L/D$$<$1. This work explores the phase properties of colloids with aspect ratios $L^*$=\,0.5, 0.3, 0.2 and 0.1. It must be noted that if an analogy with the Gay-Berne model is sought, $L^*$=$\kappa$.
 
The interaction between particles in the Kihara fluid is essentially given by a 12-6 Lennard-Jones pair potential adapted to the colloidal shape. The model follows from the seminal idea of Taro Kihara of considering a dependence of the pair interaction potential on the minimum distance between the geometrical core of the particles  \cite{KIH53}. The Kihara model has been extensively applied to the characterization of the dynamic and thermodynamic behavior and the liquid crystal phase diagrams of disk-like \cite{MAR09} and rod-like \cite{VEG91,VEG92,CUE03} colloids. Here, we extend the reach of the previous investigations to the regime of low temperatures. To this end, we employ the truncated and shifted version of Kihara the interaction potential, represented in Fig.\,\ref{fig1}:  

\begin{equation}
U(d_m) = 4\varepsilon \left[ \left( \frac{L}{d_m}\right)^{12} - \left( \frac{L}{d_m}\right)^6 -\left( \frac{L}{d_c}\right)^{12} + \left( \frac{L}{d_c}\right)^6  \right]
\end{equation}\label{eq1}	

The Kihara potential depends on the relative distance and orientation of the particles only implicitly through $d_m$=\,$d_m$({\bf r},${\bf e_i}$,${\bf e_j}$). Here, $d_m$ represents the minimum distance between the core disks of the interacting particles, which is a non-analytical function of the relative position of both particles, defined by vector ${\bf r}$, and their orientation vectors, characterize by the unitary vectors ${\bf e_i}$ and ${\bf e_j}$ as illustrated in Fig.\,\ref{fig1}. 
Efficient algorithms for the computation of $d_m$ have been reported in the literature \cite{CUE08}. Here, the interaction potential is truncated at $d_m$=\,$d_c$=\,3$L$, and consistently shifted to ensure continuity. Throughout the study, $\varepsilon$ and $D$ will be used as units of energy and length, respectively. The packing fraction is defined as $\eta$=\,$Nv_p/V$ where $N$ is the total number of particles, $v_p$ is the volume of the core spherocylinder and $V$ the total system volume \cite{BOU86}.

Two combined strategies are explored for the elucidation of the phase diagram of the colloidal fluids, namely isothermal-isobaric Monte Carlo (MC-NPT) simulations on the one hand, and direct coexistence procedures on the other hand, inspired by the work of Prof. Luis Rull in his late career \cite{RUL17a,RUL17b}. 

\subsection{MC-NPT simulations}

The MC-NPT simulations are based on Monte Carlo cycles at constant temperature $T^*$=\,$k_BT/\varepsilon$ and pressure $P^*$=\,$PD^3/\varepsilon$.  Each MC cycle involves N attempts of randomly chosen particle displacements and/or reorientations plus a trial change in the box volume. Volume changes are attempted by randomly changing the length of each side of the box independently, using the usual periodic boundary conditions. Acceptance coefficients are kept within 30-50\% for particle tilt and displacement, and 25-30\% for the box volume change. Previous studies at higher temperatures than the ones presently explored have shown that the columnar phase is the most stable liquid crystal phase for discoidal particles at sufficiently high density \cite{CUE08,MAR09,MOR21}. Based on this information, the present MC-NPT simulations were initiated from columnar arrangements packed in hexagonal arrays. Long equilibration runs corroborated that the columnar phase is stable at high pressure for all particle sizes here explored. The resulting columnar configurations were then expanded at a constant temperature in successive runs at decreasing pressures. The number of particles ranged $N\in[2500-4000]$; the smaller aspect ratios demanded the larger numbers to ensure appropriate box sizes. Typically, for each pressure $2\cdot10^6$ MC cycles were needed to equilibrate the system, and additional $5\cdot10^5$ MC cycles to obtain ensemble averages. 

The different liquid crystal phases adopted by the discotic colloids were characterized with specific order parameters and distribution functions. The orientational order was monitored using the standard nematic order parameter $S_2$ and the nematic unitary director vector $\hat{\textbf{n}}$ \cite{ALL93}. Long-range columnar or lamellar positional order was assessed by means of the directional distribution functions $g^0_{||}$, $g_{||}$, $g^0_{\perp}$ and $g_{\perp}$, describing correlations in directions parallel and perpendicular to the nematic director \cite{CUE08,MAR09,MOR21}. The structural information provided by these correlation functions is described below in greater detail.

\begin{figure*}[ht]
\centering
\hspace*{-0.3cm}
  \includegraphics[width=9cm]{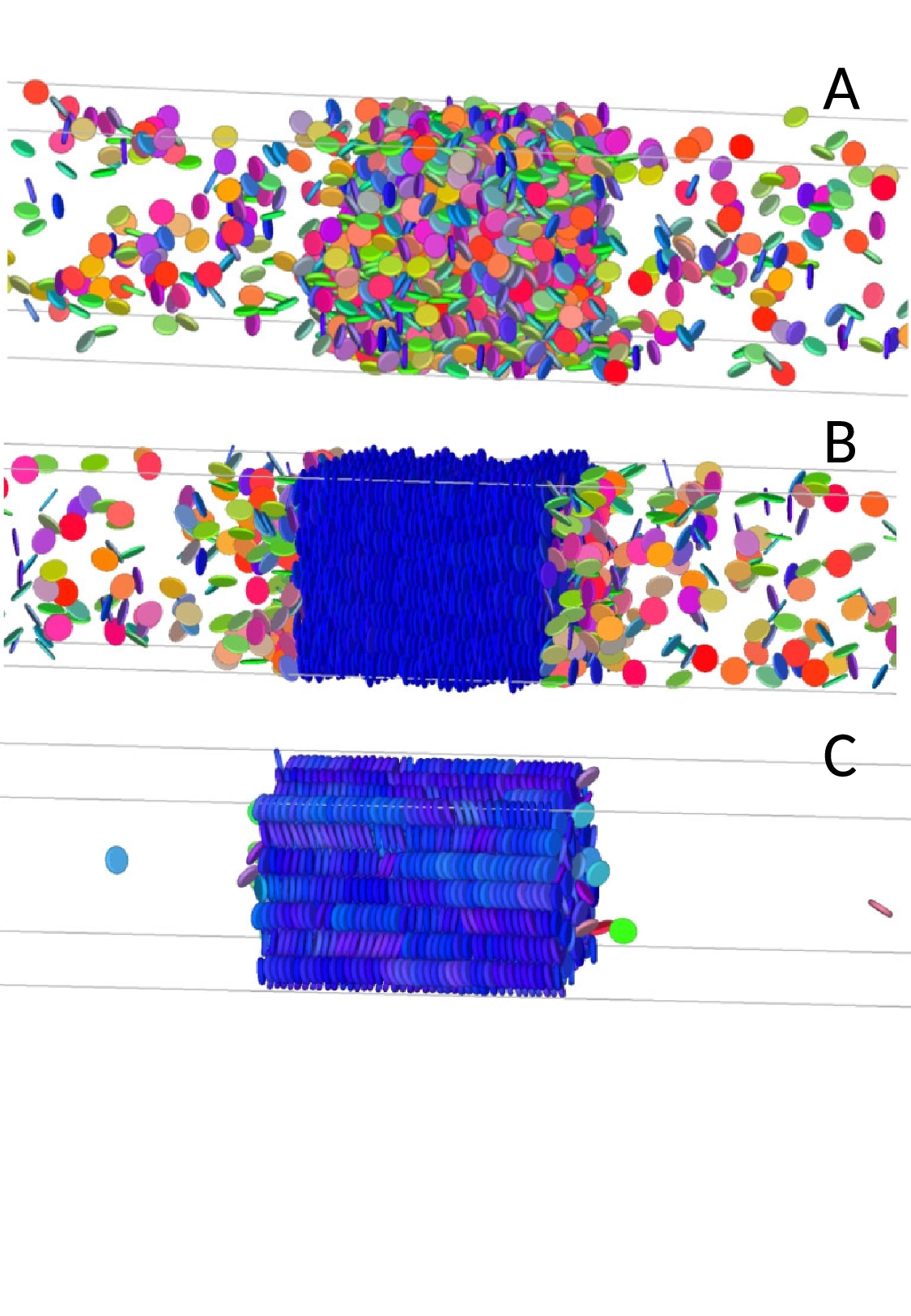}
  \vspace*{-3.cm}
  \caption{Typical simulation cells to assess vapour coexistence densities: (A) isotropic liquid $L^*$=\,0.2, $T^*$=\,0.40; (B) Lamellar liquid crystal, $L^*$=\,0.1, $T^*$=\,0.33; (C) Columnar liquid crystal, $L^*$=\,0.2, $T^*$=\,0.30. The latter panel represents a limiting case of low vapour pressure.}
  \label{fig:coex}
\end{figure*}

\subsection{Direct coexistence method}

The direct coexistence approach is based on the stabilisation of the internal structure and density of the fluid and its vapour, in which case it is assumed that an equilibrium state of coexistence has been reached. Away from coexistence, the fluid progresses towards an ideal configuration of infinite dilution. The approach is typically valid for thermodynamical states in which the coexisting phases have sufficiently different densities, preventing large volume fluctuations. Therefore this technique will not be feasible in the proximity of the critical point. 

The procedure is initiated with a $MC-NPT$ simulation at zero pressure ($P^*=0$), seeded with a configuration in the denser coexisting phase. During this simulation, the fluid progresses to a density close to the coexistence value. The resulting configuration is subsequently placed in a box with empty volumes along the positive and negative directions of a given axis. Long $NVT$ simulations (up to $5\cdot10^6$ cycles) are performed until equilibration is reached, with particles incorporating to the low-density vapour. Additional $5\cdot10^5$ $MC$-$NVT$ cycles are run to obtain ensemble averages. The densities of the liquid and vapour phases are determined from the density profile in the direction perpendicular to the interface between the two phases. Fig.\,\ref{fig:coex} illustrates typical coexistence configurations obtained for an isotropic liquid and for states within the columnar and  lamellar liquid crystal phases found in this work. In the case of an isotropic colloidal liquid, the direction in which the empty boxes are added is irrelevant.  On the contrary, for ordered phases, it is not $a$ $priori$ equivalent to add the empty boxes in the direction of the nematic director, than in a direction perpendicular to it. The results considered here were obtained by placing the empty boxes in the direction of the nematic director (see Fig.\,\ref{fig:coex}). Test runs with the empty boxes in the direction perpendicular to the nematic director led to equivalent results for the density for both phases at coexistence.

\begin{figure*}[ht]
\centering
\hspace*{-0.3cm}
  \includegraphics[width=15cm]{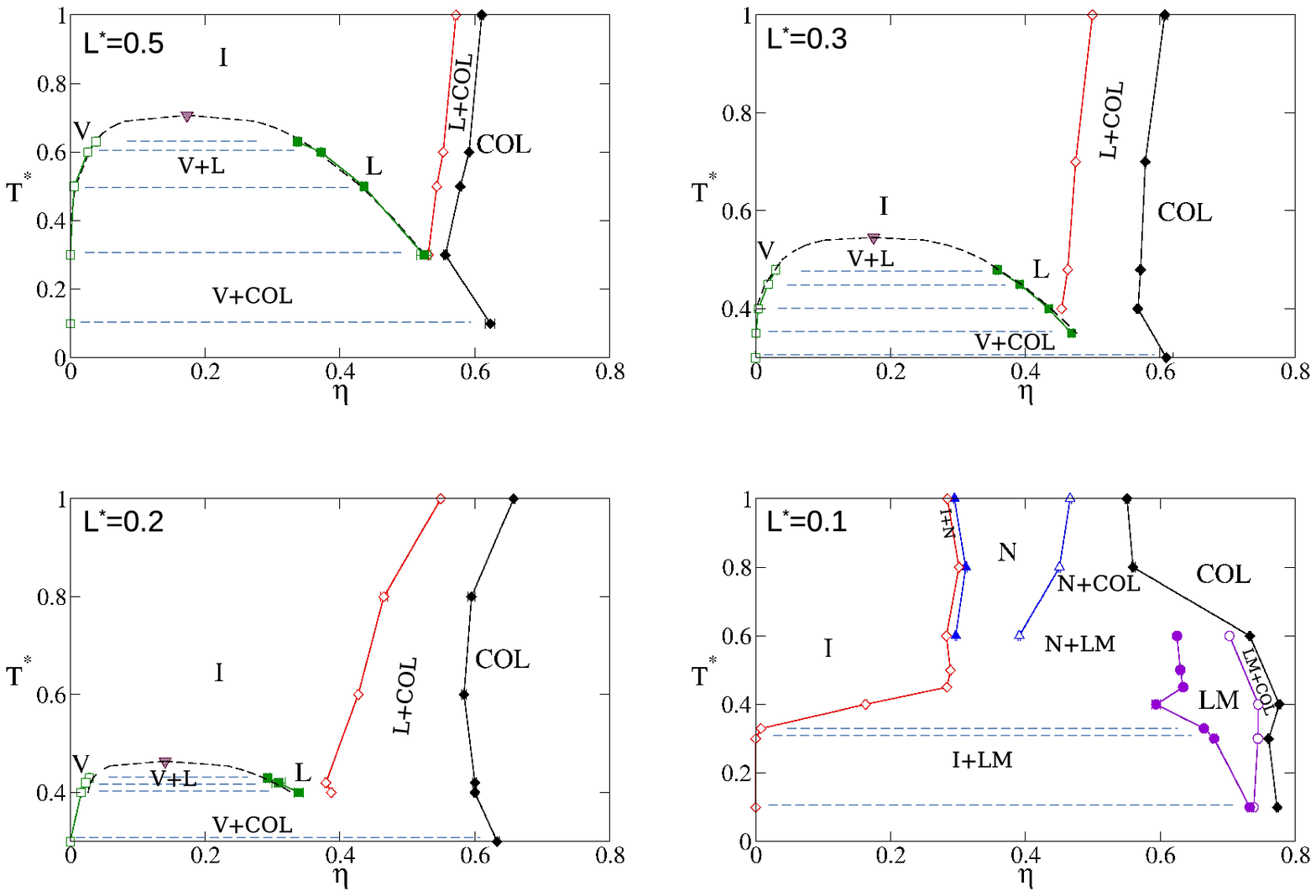}
  \vspace*{-0.4cm}
  \caption{Temperature $vs.$ packing fraction low-temperature phase diagram for Kihara fluids with $L^*$=,0.5, 0.3, 0.2 and 0.1. Red open and black solid  diamonds represent the boundaries of the isotropic liquid and the columnar phases, respectively. Green solid and open squares correspond to the packing fractions of the liquid and gaseous states in coexistence. For the case of $L^*$=\,0.1, open and solid violet circles are use to represent the high and low packing boundaries of the lamellar phase, while open and solid triangles represent the corresponding boundaries of the nematic phase. For $L^*$=\,0.5, 0.3 and 0.2, the down triangle and dashed line represent extrapolated estimates of the vapour-liquid critical point (see the text for details). In all cases, lines connecting simulation data are meant to guide the eye. Horizontal dashed lines join the points obtained by direct coexistence.}
  \label{fig:phase}
\end{figure*}

\begin{figure}[t]
\centering
\hspace*{-0.3cm}
  \includegraphics[width=14cm]{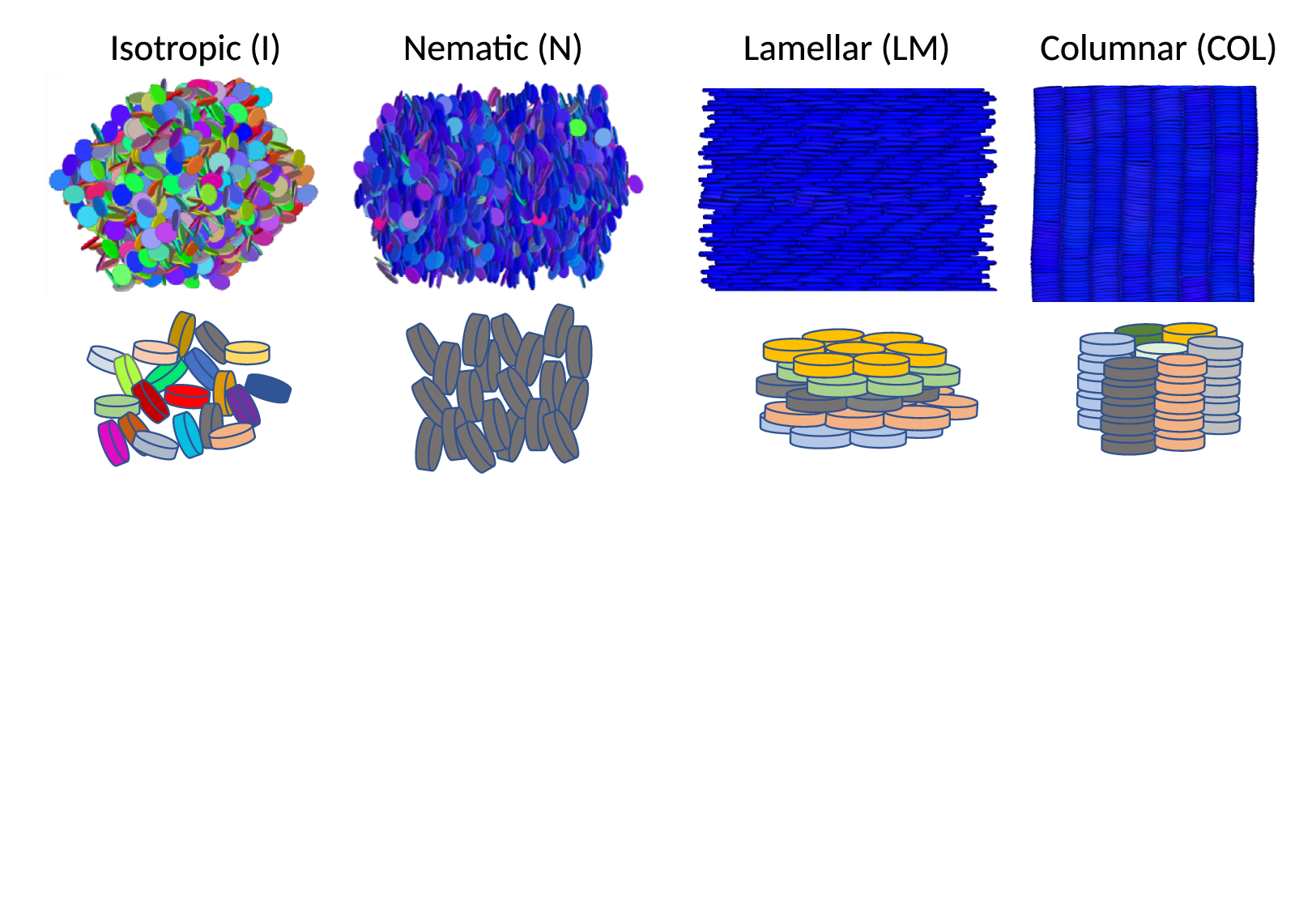}
  \vspace*{-4.8cm}
  \caption{Simulation snapshots of typical configurations and schematic representations, illustrative of the microscopic structure of the isotropic (I), nematic (N), Lamellar (LM) and Columnar (COL) Phases.}
  \label{fig:LC}
\end{figure}

\section{Results}

Vapour-liquid coexistence curves and mesoscopic liquid crystal phases were assessed for discotic Kihara colloids with aspect ratios ranging $L^*$=$L/D$=\,0.5--0.1, covering temperatures $T^*$=\,0.1--1.0. A compact representation of the phase diagrams resulting from the simulations is provided in Fig.\,\ref{fig:phase}. The mesoscopic structures of the liquid crystal phases are illustrated schematically in Fig.\,\ref{fig:LC}. 

On the one hand, the colloids display the usual isotropic ($I$), nematic ($N$) and columnar ($COL$) phases, extensively described for discotics at higher temperatures \cite{CUE08,MAR09,MAR11,MOR21}. Previous studies discerned phase boundaries for disordered, ordered and tilted columnar domains \cite{MAR09,MAR11}, but this is not attempted within the scope of this work.  The isotropic and columnar phases are present for the larger particle aspect ratios, $L^*$$\ge$\,0.2. The discotic nematic phase enters the phase diagram in the higher temperature range, $T^*$$>$\,0.5, of the $L^*$=\,0.1 fluid. The limitation of the occurrence of the nematic phase for discotic fluids of low anisotropy has been studied in depth previously for the case of hard particles \cite{CUE08,MAR11}. On the other hand, an apparently novel layered liquid crystal phase is observed with structural characteristics of a dense lamellar material (henceforth denoted $LM$), that is described below in detail. The lamellar phase emerges in the lower temperature range, $T^*$$>$\,0.1--0.6, only for the $L^*$=\,0.1 fluid.

The formation of thermodynamically stable layered arrangements of discotic particles is appealing both from a fundamental point of view, and also from the potential applications in materials design. Such phases are actually elusive, due to the typically higher stability of the columnar phases. There is limited evidence for the experimental realization of layered liquid crystalline phases in discotic particle systems. Examples of this are systems with high polidispersity \cite{SUN09}, or fluids of charged colloidal nanosheets \cite{DAV18}. Previous free-energy calculations showed that the phase diagram of the fluid of hard discotic particles with $L^*$$\ge$\,0.1 does not display any lamellar phase, suggesting that steric driving forces intrinsically favor columnar arrangements \cite{MAR11}. A reduced smectic domain was found in Monte Carlo simulations of the phase diagram of discotics with strong edge-edge interactions  \cite{MOR21}. In that case, the fluid self-organized in weakly-interacting layers, each of which behaved effectively as a two-dimensional disordered fluid, with pair correlations modulated by the strength of the edge-edge interactions. The present lamellar phase, $LM$, is qualitatively different, as it features intralayer and interlayer interactions of similar magnitude and high packing fractions within layers, as described below in detail.

\begin{figure*}[t]
	\centering
	\hspace*{-0.3cm}
	\includegraphics[width=14cm]{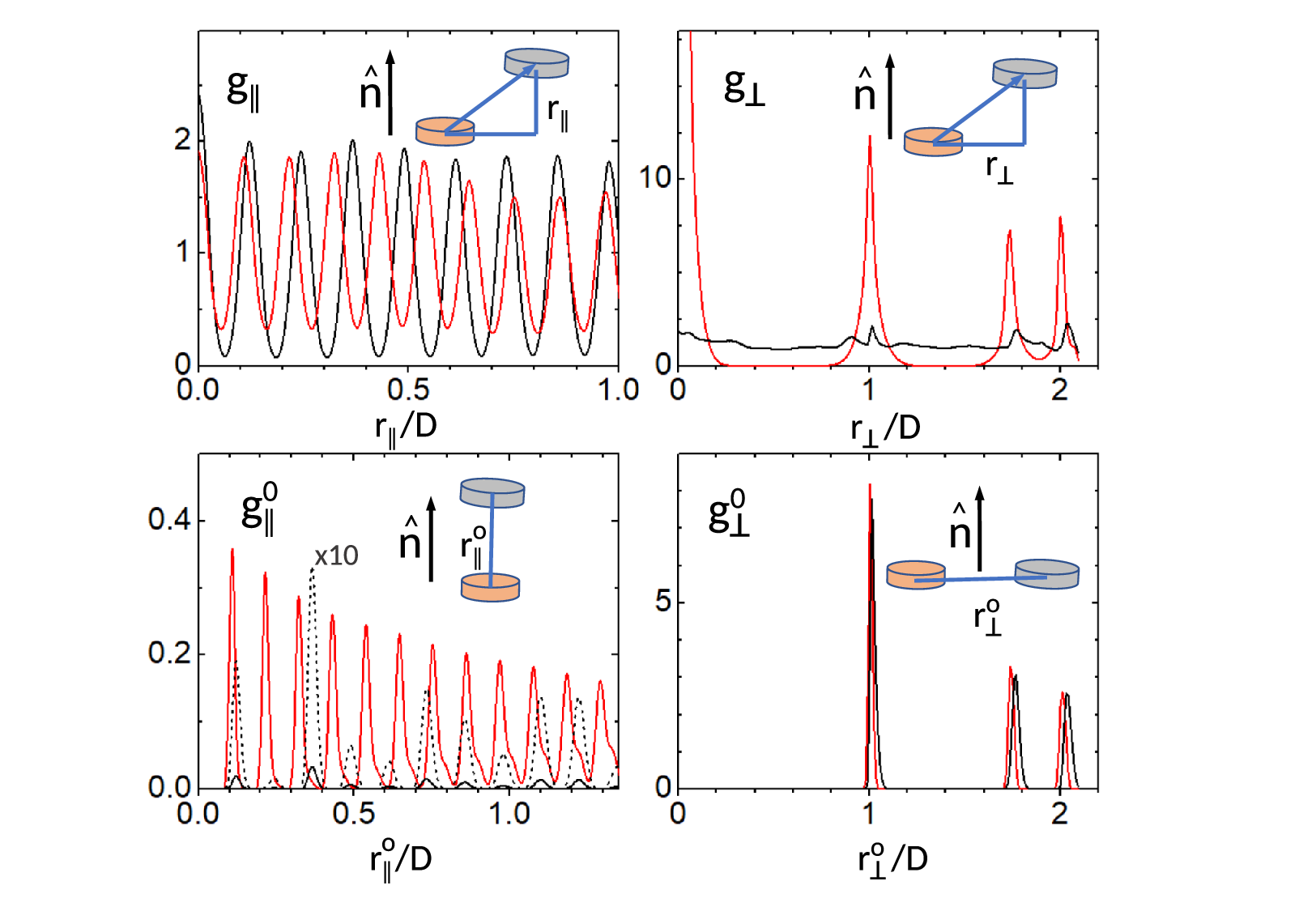}
	\vspace*{-0.4cm}
	\caption{Correlation functions $g_{||}$, $g^0_{||}$,$g_{\perp}$, $g^0_{\perp}$, characterizing the internal ordering of discotic colloids in stateswithin the lamellar ($L^*=0.1$, $\eta=0.680$, black traces) and columnar ($L^*=0.1$, $\eta=0.798$, red traces) phases. The inset of each panel provides a schematic representation of the associated coordinate guiding the correlation in each case. See text for a detailed description. The black dashed line for $g^0_{||}$ corresponds to a ten-fold magnification of the trace for the lamellar state, for a better visualization.}
	\label{fig:g}
\end{figure*}

The internal structure of the isotropic, nematic, lamellar and columnar phases was characterized by means of the $g^0_{||}$, $g_{||}$, $g^0_{\perp}$ and $g_{\perp}$ correlation functions. The information provided by the structure of these functions is illustrated in Fig.\,\ref{fig:g} for typical lamellar and columnar arrangements. The observations can be summarized as follows: 

i) $g^0_{||}(\hat{r}_{||})$ monitors particles located at a distance ${r^0}_{||}$ along the direction of the nematic director $\hat{\textbf{n}}$.
In the columnar phase, it displays a peak structure with a rough spacing of $L$, naturally reflecting the face-to-face stacking of the particles. This correlation changes qualitatively in the lamellar phase, which displays much weaker correlations (note the ten-fold magnification of the curve in Fig.\,\ref{fig:g}), with a dominant peak recurrence at a spacing of about 3$L$, indicating that the particles in neighbouring layers are displaced from each other by roughly one third of the particle diameter (see Fig.\ref{fig:LC}).  
 
ii) $g_{||}(r_{||})$ includes all particles located in a plane at a given distance along the nematic axis, hence $r_{||}$ represents the projection of the distance vector on the nematic director, $\hat{\textbf{n}}$.
The structure in $g_{||}$ exposes the organization of particles in layers. In the lamellar phase, it consistently displays a long-range array of equally spaced maxima. The columnar phase also shows a net layer correlation due to the trend of neighbouring columns to interdigitate, which induces coplanarity of particles from half of the neighboring columns \cite{CUE08}.

iii) $g^0_{\perp}(\hat{r}_{\perp})$ refers to neighboring particles at distances ${r^0}_{\perp}$ within the same plane perpendicular to the nematic director; it hence exposes the two-dimensional organization within the layers. The peak structure $g^0_{\perp}$ is similar in the lamellar and columnar arrangements of this study and is representative of a marked hexatic packing of the particles.  In the lamellar phase, this is related to the high packing fraction of each layer. In the columnar phase, it reflects the neat hexagonal packing of neighbouring columns. 

iv) $g_{\perp}(r_{\perp})$ spans particles with a given projection of the distance vector on the plane perpendicular to the nematic director, $(r_{\perp})$. This effectively includes all neighboring particles located at the surface of a cylinder of radius $(r_{\perp})$. This function effectively probes the existence of columns at given distances perpendicular to the nematic director; it hence lacks structure in the lamellar phase.

From the above description, it becomes apparent that the $LM$ phase displays a layered organization of the particles with a two-dimensional hexagonal packing within each layer. Moreover, the particles of any given layer are displaced with respect to those of the neighboring layers, by roughly one third of their diameter. The driving force for such displacement is a maximization in the number of interacting neighbours per particle with respect to the columnar arrangement. At the low temperatures in which the lamellar phase is observed, the favourable energetic factor compensates for the decrease in positional entropy associated with the constrains imposed on the out-layer translational degrees of freedom of the particles. 
Fig.\,\ref{fig:phase} shows that the lamellar $LM$ phase is stable in the $L^*$=0.1 fluid over a specific range of temperatures $T^*$$\approx$\,0.1--0.6. 
The simulation at $T^*$=\,0.6 displays the full sequence of phase transitions $COL-LM-N-I$ upon isothermal phase expansion. The vanishing of the $LM$ phase and the reentrance of the columnar phase at low temperature $T^*$$<$\,0.1 was not necessarily expected. A detailed understanding of the balance of entropic and energetic contributions that sustain the lamellar phase would plausibly require free-energy assessments similar to the ones applied to the columnar phases of discotic colloids \cite{MAR11}. It must be noted that, despite the particularly long MC runs that were applied to monitor the stability of the lamellar phase (above $5\cdot10^6$ cycles in some cases), truly homogeneous configurations were not achieved, as domains of local columnar order remained within the simulation box. Nevertheless, the dominant lamellar arrangement seemed robust, with no evidence of metastability ($e.g.$ a transient configuration for the melting of the columnar phase into a less ordered nematic or isotropic phase).

\begin{figure*}[t]
	\centering
	\hspace*{-0.3cm}
	\includegraphics[width=14cm]{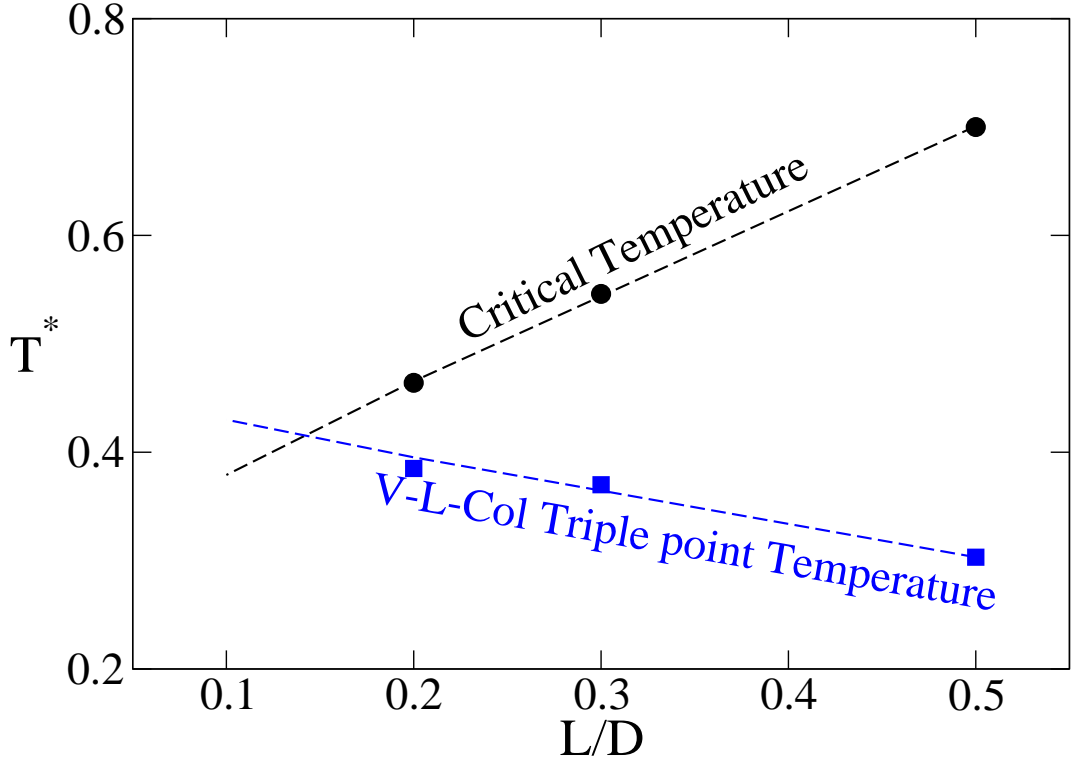}
	\vspace*{-0.4cm}
	\caption{Critical vapour-liquid (black circles) and vapour-liquid-columnar triple point (blue squares) temperatures estimated from the data shown in Fig.\,\ref{fig:phase}, as a function of particle aspect ratio, $L^*=L/D$. The dashed lines are the linear fit to the data in each case.}
	\label{fig6}
\end{figure*}

We now look in greater detail into the domains of liquid-vapour coexistence, that overlap with the liquid crystal phase behavior of the fluids under study. 
Fig.\,\ref{fig:phase} show that a 'conventional' coexistence between an isotropic liquid and its vapour is observed in the higher temperature range,  at temperatures above $T^*$$\approx$\,0.30, 0.35 and 0.40 for the $L^*$=\,0.5, 0.3 and 0.2 fluids, respectively. The liquid-vapour coexistence domain narrows down rapidly as the critical temperature T$^*_c$ decreases with increasing particle anisotropy. The critical temperatures may be estimated from a fit of the simulation data to the law of rectilinear diameters, considering a critical exponent of $\beta=0.32$ \cite{FrenkelSmit}

\begin{equation}
\eta_l(T^*)+\eta_v(T^*) = 2\eta_c+A(T^*-T^*_c)
\end{equation}\label{eq-c1}	
\begin{equation}
\eta_l(T^*)-\eta_v(T^*) = B(T^*-T^*_c)^{\beta}
\end{equation}\label{eq-c2}	

Assuming the simultaneous validity of these expressions leads to  $T^*_c$$\sim$0.70, 0.55 and 0.46 for $L^*$=\,0.5, 0.3 and 0.2, respectively, whereas no critical point is found for $L^*$=\,$0.1$. It can be estimated at what value of $L^*$ the vapour-liquid coexistence will disappear by estimating at what size the critical temperature curve as a function of size will cross the vapour-liquid-columnar triple point temperature curve as a function of size. From the coexistence curves in Fig.\,\ref{fig:phase} we estimate the temperature of these triple points to be $T_{tp}^* \sim 0.3, 0.37$ and $0.385$ for $L^*=0.5, 0.3$ and $0.2$,  respectively. In Fig.\,\ref{fig6}, $T_c^*$ and $T_{tp}$* are plotted against $L^*$, showing that the linear extrapolations of the two sets of temperature values cross at $L^*\sim 0.145$, a size that can be taken as an estimate of the disappearance of vapour-liquid coexistence, although this value  should be handled with care, due to the small number of points, and the uncertainties in the estimation of the triple point temperatures.

This coexistence regime connects with a previous study based on Gibbs ensemble Monte Carlo (GEMC) simulations for Kihara discotics with $L^*\ge 0.5$ \cite{Gam08}. The present coexistence densities for the $L^*$=\,0.5 fluid ovelap nicely with the GEMC results for the same fluid (the critical temperature derived from the GEMC data is slightly higher, T$^*_c$$\sim$0.75, but within statistical uncertainty). 

The liquid crystal phases incorporate to the vapour-liquid coexistence at the lower temperatures. The application of the GEMC method is not feasible in this region due to inherent inefficiencies of the insertion method. For the $L^*$=\,0.5, 0.3, 0.2 and 0.1 colloids, columnar-vapour equilibrium is observed at $T^*$$<$\,0.30, 0.35, 0.40 and 0.60, respectively. The $L^*$=\,0.1 fluid actually represents a limiting situation of high particle anisotropy in which no isotropic liquid-vapour coexistence is observed at any temperature. The coexistence of the nematic phase with the isotropic phase resembles that of a conventional liquid crystal transition with a moderate change in density. The lamellar phase, and eventually also the columnar phase, do display vapour coexistence. The density of the vapour branch in such coexistence is however challenging to assess due to the largely reduced vapour pressure and the correspondingly low statistical significance of the sampling of the vapour phase. A limiting case of low vapour pressure is illustrated in Fig.\,\ref{fig:coex} by the simulation cell for the columnar state of the $L^*$=\,0.2 system at $T^*$=\,0.30. Despite these limitations in the low temperature limit, the coexistence cell approach has allowed to assess domains of vapour equilibrium for liquid crystal phases with packing fractions far beyond the reach of Gibbs ensemble Monte Carlo methods.

\section{Conclusions}

Considering the extensive body of knowledge built in the past decades on the liquid crystalline phase behaviour of discotic particle fluids, studies focusing on the phase behaviour at low temperatures have been scarce, with little evidence on vapour-liquid coexistence. With this study we aim to contribute to fill this gap, and gain novel insights into the phase behaviour of oblate particle fluids.

One main contribution of this work was been the elucidation of vapour-liquid coexistence for fluids with highly asymmetric particles under high density regimes. This feature has been challenging for Monte Carlo approaches due to the reduced particle insertion rates, which allowed to model the isotropic vapour-liquid coexistence for particles with moderate aspect ratios \cite{Gam08}. One additional difficulty relates to the fact, corroborated here, that the critical point temperature decreases rapidly as the particle anisotropy increases, eventually merging the vapour-liquid coexistence regime with the liquid crystal diagram of the system. Following this trend, we find that the isotropic vapour--liquid coexistence vanishes for $L^*$=\,0.1, where it is replaced by liquid crystal--vapour coexistence. 

Particle anisotropy emerges as a crucial parameter in discoidal fluids. At low and moderate aspect rations, $L^*\ge$\,0.2, the discotic phase diagram maintains a qualitatively similar topology, involving vapour-liquid coexistence and vapour-columnar coexistence in the low temperature limit \cite{CUE08,MAR11}. At high anisotropy, $L^*=$\,0.1, the nematic phase emerges, along with further ordered phases. To this respect, a remarkable result of this study has been the finding of a novel phase for the discotic particles with a lamellar structure. In this phase, the particles are arranged in layers perpendicular to the nematic director. Within these layers there is a high level of two-dimensional hexagonal order, and the lattice of one layer is displaced with respect to the adjacent layers, thus disrupting the common columnar structure adopted by discotics at high density. Computer simulations suggest that strong interactions are needed to stabilize the lamellar phase \cite{MOR21,DAV18}, but the stability of the lamellar phase may be enhanced by the existence of polydispersity in the dimensions of the particles \cite{SUN09}. The combination of both effects could extend the range of existence of this elusive phase. We expect this to be the subject of future research. This lamellar phase described joins the plethora of phases reported in other works for discotic particle fluids \cite{CUE08,MAR11,MOR21}, which turns out to be richer than in the case of elongated particle fluids.

It should be stressed that the results presented here correspond to situations where the interaction between the particles maintains a spherocylindrical symmetry, with no specific dependence on the relative orientation of the particles. Recently, our group has reported a catalogue of phases and fluid structures at high temperature for discotics with interactions explicitly dependent on the relative orientation of the particles \cite{MOR21}. In the light of the results of this work, it is foreseen  that the phase diagram of discotics may be further enriched at low temperatures by the introduction of orientational interactions. Their effect on vapour and liquid crystal coexistence at low temperature remains an open question that we intend to address in the future, thus continuing the work and legacy of the much missed Prof. Luis F. Rull.\\

{\bf Acknowledgements}\\
The authors are deeply grateful to prof. Luis F. Rull for his  scientific guidance, continuous inspiration and friendship. At a much less important level, we acknowledge financial support from the Ministry of Science of Spain and FEDER (projects TED2021-130683B-C21, PID2021-126121NB-I00 and PID2021-126348NB-I00). We are grateful to C3UPO of Universidad Pablo de Olavide for High-Performace Computing support.

\bibliographystyle{tfo}
\bibliography{kihara}

\begin{thebibliography}{48}
\providecommand{\url}[1]{\texttt{#1}}
\providecommand{\urlprefix}{URL }

\bibitem{DeGennes1993}
P.G. De~Gennes and J. Prost, \emph{The Physics of Liquid Crystals}, 2nd ed.
  (Clarendon Press, Oxford, United Kingdom, 2008).

\bibitem{deMiguel92}
E. De~Miguel and M.P. Allen,  Molecular Physics  \textbf{76} (6), 1275--1279
  (1992).

\bibitem{VEG92}
C. Vega, S. Lago, E. De~Miguel and L.F. Rull,  The Journal of Physical
  Chemistry  \textbf{96} (18), 7431--7437 (1992).

\bibitem{deMiguel1}
E. de~Miguel, E.M. del R\'{\i}o, E.T. Brown and M.P. Allen,  Journal of
  Chemical Physics  \textbf{105}, 4234--4249 (1996).

\bibitem{WU}
L. Wu, G. Jackson and E. M\"uller,  International Journal of Molecular Sciences
   \textbf{14}, 16414--16442 (2013).

\bibitem{Levesque}
D. Levesque, J.J. Weis and G.J. Zarragoicoechea,  Physical Review E
  \textbf{47} (1), 496--505 (1993).

\bibitem{Williamson}
D.C. Williamson and F. del Rio,  Journal of Chemical Physics  \textbf{107}
  (22), 9549--9558 (1997).

\bibitem{McGrother}
S.C. McGrother, A. Gil-Villegas and G. Jackson,  Molecular Physics  \textbf{95}
  (3), 657--673 (1998).

\bibitem{Houssa}
M. Houssa, A. Oualid and L.F. Rull,  Molecular Physics  \textbf{94} (3),
  439--446 (1998).

\bibitem{Houssa2}
M. Houssa, L.F. Rull and S.C. McGrother,  Journal of Chemical Physics
  \textbf{109} (21), 9529--9542 (1998).

\bibitem{Williamson2}
D.C. Williamson, N.A. Thacker and S.R. Williams,  Physical Review E
  \textbf{71} (2, 1) (2005).

\bibitem{Houssa3}
M. Houssa, L.F. Rull and J.M. Romero-Enrique,  Journal of Chemical Physics
  \textbf{130} (15) (2009).

\bibitem{Tre23}
V. Trejos, F. G\'amez and B. Garz\'on,  Journal of Molecular Liquids
  \textbf{383}, 122177 1--12 (2023).

\bibitem{BIS10}
H.K. Bisoyi and S. Kumar,  Chem. Soc. Rev.  \textbf{39}, 264--285 (2010).

\bibitem{WOH16}
T. Wohrle, I. Wurzbach, J. Kirres, A. Kostidou, N. Kapernaum, J. Litterscheidt,
  J.C. Haenle, P. Staffeld, A. Baro, F. Giesselmann and S. Laschat,  Chemical
  Reviews  \textbf{116} (3), 1139--1241 (2016).

\bibitem{Brown2}
A.B.D. Brown, S.M. Clarke and A.R. Rennie,  Langmuir  \textbf{14} (11),
  3129--3132 (1998).

\bibitem{vanderKooij}
F.M. van~der Kooij, D. van~der Beek and H.N.W. Lekkerkerker,  Journal of
  Physical Chemistry B  \textbf{105} (9), 1696--1700 (2001).

\bibitem{vanderKooij2}
F.M. van~der Kooij, K. Kassapidou and H.N.W. Lekkerkerker,  Nature
  \textbf{406} (6798), 868--871 (2000).

\bibitem{vanderKooij3}
F.M. van~der Kooij and H.N.W. Lekkerkerker,  Journal of Physical Chemistry B
  \textbf{102} (40), 7829--7832 (1998).

\bibitem{Liu}
S.Y. Liu, J. Zhang, N. Wang, W.R. Liu, C.G. Zhang and D.J. Sun,  Chemistry of
  Materials  \textbf{15} (17), 3240--3241 (2003).

\bibitem{Fossum}
J.O. Fossum, E. Gudding, D.D.M. Fonseca, Y. Meheust, E. DiMasi, T. Gog and C.
  Venkataraman,  Energy  \textbf{30} (6), 873--883 (2005), Conference on
  Transport, Dissipation and Vortices, Trondheim, Norway, JUN 01-05, 2003.

\bibitem{Michot}
L.J. Michot, I. Bihannic, S. Maddi, S.S. Funari, C. Baravian, P. Levitz and P.
  Davidson,  Proceedings of the National Academy of Sciences of the United
  States of America  \textbf{103} (44), 16101--16104 (2006).

\bibitem{Veerman}
J.A.C. Veerman and D. Frenkel,  Physical Review A  \textbf{45} (8), 5632--5648
  (1992).

\bibitem{GB}
J.G. Gay and B.J. Berne,  J.~Chem. Phys.  \textbf{74} (6), 3316--3319 (1981).

\bibitem{KIH53}
T. Kihara,  Reviews of Modern Physics  \textbf{25} (4), 831--840 (1953).

\bibitem{Gam08}
F. Gámez, S. Lago, B. Garzón, P.J. Merkling and C. Vega,  Molecular Physics
  \textbf{106} (11), 1331--1339 (2008).

\bibitem{MAR09}
B. Martínez-Haya and A. Cuetos,  The Journal of Chemical Physics  \textbf{131}
  (7), 074901 (2009).

\bibitem{MOR21}
N. Morillo, B. Martínez-Haya and A. Cuetos,  Soft Matter  \textbf{17},
  8693--8704 (2021).

\bibitem{deMiguel}
E. De~Miguel, L. Rull, M. Chalam and K. Gubbins,  Molecular Physics
  \textbf{71} (6), 1223--1231 (1990).

\bibitem{Elvira}
E. Del~Rio, E. De~Miguel and L. Rull,  Physica A  \textbf{213} (1-2), 138--147
  (1995).

\bibitem{Brown}
E.T. Brown, M.P. Allen, E.M. del R\'{\i}o and E. de~Miguel,  Physical Review E
  \textbf{57}, 6685--6699 (1998).

\bibitem{Emerson2}
A.P.J. Emerson, G.R. Luckhurst and S.G. Whatling,  Molecular Physics
  \textbf{82}, 113--124 (1994).

\bibitem{Bates4}
M.A. Bates and G.R. Luckhurst,  Journal of Chemical Physics  \textbf{104},
  6696--6709 (1996).

\bibitem{Ryckaert1}
D. Caprioni, L. Bellier-Castella and J.P. Ryckaert,  Physical Review E
  \textbf{67}, 041703 (2003).

\bibitem{Ryckaert2}
L. Bellier-Castella, D. Caprioni and J.P. Ryckaert,  Journal of Chemical
  Physics  \textbf{121}, 4874--4883 (2004).

\bibitem{Chakrabarti}
D. Chakrabarti and D.J. Wales,  Physical Review E  \textbf{77}, 051709 (2008).

\bibitem{Frenkel}
D. Frenkel and B.M. Mulder,  Molecular Physics  \textbf{55}, 1171--1192 (1985).

\bibitem{RUL17a}
L.F. Rull and J.M. Romero-Enrique,  Molecular Physics  \textbf{115} (9-12),
  1214--1224 (2017).

\bibitem{RUL17b}
L. Rull and R.E. J.M.,  Langmuir  \textbf{33}, 11779--11787 (2017).

\bibitem{VEG91}
C. Vega and S. Lago,  Chemical Physics Letters  \textbf{185} (5), 516--521
  (1991).

\bibitem{CUE03}
A. Cuetos, B. Mart\'{\i}nez-Haya, S. Lago and L.F. Rull,  Phys. Rev. E
  \textbf{68}, 011704 (2003).

\bibitem{BOU86}
T. Boubl\'{\i}k and I. Nezbeda,  Collection of Czechoslovak Chemical
  Communications  \textbf{51}, 2301--2432 (1986).

\bibitem{CUE08}
A. Cuetos and B. Mart{\'{i}}nez-Haya,  The Journal of Chemical Physics
  \textbf{129} (21), 214706 (2008).

\bibitem{ALL93}
M.P. Allen, G.T. Evans, D. Frenkel and B.M. Mulder, in \emph{Advances in
  chemical physics}, Vol.~86  (, , 2007), pp. 1--166.

\bibitem{MAR11}
M. Marechal, A. Cuetos, B. Martínez-Haya and M. Dijkstra,  Journal of Chemical
  Physics  \textbf{134}, 094501 1--8 (2011).

\bibitem{SUN09}
D. Sun, H.J. Sue, Z. Cheng, Y. Martínez-Ratón and E. Velasco,  Physical
  Review E  \textbf{80}, 041704 (2009).

\bibitem{DAV18}
P. Davidson, C. Penisson, D. Constantin and J.C.P. Gabriel,  Proceedings of the
  National Academy of Sciences  \textbf{115} (26), 6662--6667 (2018).

\bibitem{FrenkelSmit}
D. Frenkel and B. Smit, \emph{Understanding Molecular Simulation: From
  Algorithms to Applications}, \emph{Computational Science Series}, Vol.~1, 2nd
  ed.   (Academic Press, San Diego, 2002).

\end{thebibliography}

\end{document}